\documentclass[twocolumn]{revtex4}
\usepackage{graphicx}

\def\duzomniejsze{<\kern-.7mm<}
\def\duzowieksze{>\kern-.7mm>}

\def\textbf#1{{\bf #1}}
\def\beq{\begin{equation}}
\def\eeq{\end{equation}}
\def\be{\begin{equation}}
\def\ee{\end{equation}}
\def\ben{\begin{eqnarray}}
\def\een{\end{eqnarray}}
\def\beqa{\begin{eqnarray}}
\def\eeqa{\end{eqnarray}}
\def\eea{\end{array}}
\def\bea{\begin{array}}
\newcommand{\bei}{\begin{itemize}}
\newcommand{\eei}{\end{itemize}}
\newcommand{\bee}{\begin{enumerate}}
\newcommand{\eee}{\end{enumerate}}

\def\>{\rangle}
\def\<{\langle}

\begin{document}

\title{Stationary state entanglement and total correlation of two qubits or qutrits}

\begin{abstract}
We investigate the mutual information and entanglement of
stationary state of two locally driven qubits under the influence
of collective dephasing. It is shown that both the mutual
information and the entanglement of two qubits in the stationary
state exhibit damped oscillation with the scaled action time
$\gamma{T}$ of the local external driving field. It means that we
can control both the entanglement and total correlation of the
stationary state of two qubits by adjusting the action time of the
driving field. We also consider the influence of collective
dephasing on entanglement of two qutrits and obtain the sufficient
condition that the stationary state is entangled.

PACS numbers: 03.65.Ud, 03.67.-a, 05.40.Ca
\end{abstract}
\author{Shang-Bin Li}\email{sbli@zju.edu.cn},\author{Jing-Bo Xu}

\affiliation{Chinese Center of Advanced Science and Technology
(World Laboratory), P.O.Box 8730, Beijing, People's Republic of
China;} \affiliation{Zhejiang Institute of Modern Physics and
Department of Physics, Zhejiang University, Hangzhou 310027,
People's Republic of China}

\maketitle

\section * {I. INTRODUCTION}

Quantum entanglement plays a very important role in quantum
information processes, which can exhibit the nature of a
nonclassical correlation between quantum systems that have no
classical interpretation \cite{Nielsen2000}. Entanglement of two
or more subsystems can be destroyed by the interaction between
quantum systems of interest and its surrounding environments.
Certain kind of the interaction between the physical system and
environments can lead to the collective dephasing, which occurs in
various physical systems such as two coupled quantum dots.
Recently, the quantum information processes in the presence of the
collective dephasing have intrigued much attention
\cite{Khodjasteh2002,Yu2002,Yu2003,Hill2004}. Khodjasteh and Lidar
have investigated the universal fault-tolerant quantum computation
in the presence of spontaneous emission and collective dephasing
\cite{Khodjasteh2002}. Hill and Goan have studied the effect of
dephasing on proposed quantum gates for the solid-state Kane
quantum computing architecture \cite{Hill2004}. Collective
dephasing allows the existence of the so-called decoherence-free
subspace \cite{Zanardi2001}. It has been shown that some kinds of
entangled states of two qubits are very fragile in the presence of
collective dephasing while the others are very robust
\cite{Yu2002}. So, it is very desirable to protect the fragile
entangled states from completely losing their entanglement in the
collective dephasing.

The mutual information is also a very important quantity which has
been used to measure the total correlation of two subsystems. In
this paper, we comparatively investigate the mutual information
measuring the total correlation and the stationary state
entanglement quantified by concurrence of two locally driven
qubits in the presence of collective dephasing. It is shown that
one can transform the fragile entangled states into the stationary
entangled states under the collective dephasing by making use of a
finite-time external driving field. We further find that the
stationary state achieving the local maximal value of concurrence
also have the local maximal value of mutual information. Moreover,
we investigate the model of two spin 1 in the presence of
collective dephasing and obtain the general sufficient condition
that the stationary state is entangled.

The disentanglement of entangled states of qubits is a very
important issue for quantum information processes, such as the
solid state quantum computation. For example, in quantum
registers, some kinds of undesirable entanglement between the
qubits can lead to the decoherence of the qubit \cite{Reina2002}.
Recently, Yu and Eberly have found that the time for decay of the
qubit entanglement can be significantly shorter than the time for
local dephasing of the individual qubits \cite{Yu2002,Yu2003}. The
collective dephasing can be described by the master equation \be
\frac{\partial\hat{\rho}}{\partial{t}}=\frac{\gamma}{2}(2\hat{J}_{z}\hat{\rho}\hat{J}_{z}-\hat{J}^2_{z}\hat{\rho}-\hat{\rho}\hat{J}^2_{z}),
\ee where $\gamma$ is the decay rate. $\hat{J}_{z}$ are the
collective spin operator defined by \be
\hat{J}_{z}=\sum^{2}_{i=1}\hat{\sigma}^{(i)}_{z}/2,\ee where
$\hat{\sigma}_z$ for each qubit is defined by
$\hat{\sigma}_{z}=|1\rangle\langle{1}|-|0\rangle\langle{0}|$.
Previous studies have shown that two of the four Bell states
$|\Psi^{\pm}\rangle\equiv\frac{\sqrt{2}}{2}(|11\rangle\pm|00\rangle)$
are fragile states in the collective dephasing channel, while the
others Bell states
$|\Phi^{\pm}\rangle\equiv\frac{\sqrt{2}}{2}(|10\rangle\pm|01\rangle)$
are robust entangled states.

Meanwhile, much attention has been paid to the quantum information
processes in which the basic element is a qutrit. Quantum
communication complexity protocol with two entangled qutrits has
been proposed \cite{Brukner2002}, and it has been proven that, for
a broad class of protocols the entangled state of two qutrits can
enhance the efficiency of solving the problem in the quantum
protocol over any classical one if and only if the state violates
Bell's inequality for two qutrits. By making use of the entangled
qutrits, a generalization of Ekert's entanglement-based quantum
cryptographic protocol has been discussed \cite{Durt2003}. By
means of the orbital angular momentum of the photons, the
experimental implementation of producing, transmitting, and
reconstructing the qutrit has been reported
\cite{Molina-Terriza2004}. The generation, manipulation and
measurement of entangled qutrits have also been experimental
studied by utilizing spontaneous parametric down converted photons
and unbalanced 3-arm fiber optic interferometers \cite{Thew2004}.
In ideal situations, entangled qutrits provide better security
than qubits in quantum bit commitment. However, it has been shown
that qutrits with even a small amount of decoherence cannot offer
increased security over qubits \cite{Langford2004}. It motivates
us to investigate how the decoherence can affect the entanglement
of two qutrits. We consider two qutrits in collective dephasing
channel and obtain the sufficient condition for the entangled
stationary state. By making use of this sufficient condition, we
can distinguish the robust entangled state of two qutrits from the
fragile entangled state.

This paper is organized as follows: In section II, we
comparatively investigate the mutual information measuring the
total correlation and the stationary state entanglement quantified
by concurrence of two locally driven qubits in the presence of
collective dephasing. In section III, we consider two qutrits in
collective dephasing channel and give a sufficient criterion that
a class of initial entangled state of two qutrits which is robust
against the collective dephasing. In section IV, there are some
conclusions.

\section * {II. THE ENTANGLEMENT AND MUTUAL INFORMATION OF THE STATIONARY STATE}

In this section, we investigate the model in which two qubits are
exposed in a collective dephasing environment and one of two
qubits is simultaneously driven by a finite-time external field.
The dynamics of two qubits can be described by the following
master equation \be
\frac{\partial\hat{\rho}}{\partial{t}}=-\frac{i}{2}[\Omega_1(t)\hat{\sigma}^{(1)}_x,\hat{\rho}]+\frac{\gamma}{2}(2\hat{J}_{z}\hat{\rho}\hat{J}_{z}-\hat{J}^2_{z}\hat{\rho}-\hat{\rho}\hat{J}^2_{z}),
\ee where $\Omega_{1}(t)=\Omega_1\Theta(T-t)$ is the intensity of
the time-dependent external driving field acted on the qubit 1,
and $\Theta(x)$ is the unit step function and equals one for
$x\geq0$ and equals zero for $x<0$. In the following calculations,
we will show that the action time $T$ of the external driving
field plays a special role in the stationary state entanglement
and the total correlation of two qubits. After numerically solving
the master equation (3), we can obtain the stationary state
density matrix of two qubits. Without loss of generality, the
stationary state density matrix has the form \beqa
\rho_s&=&a(\frac{\Omega_1}{\gamma},\gamma{T})|11\rangle\langle11|+b(\frac{\Omega_1}{\gamma},\gamma{T})|10\rangle\langle10|\nonumber\\
&&+c(\frac{\Omega_1}{\gamma},\gamma{T})|01\rangle\langle01|+d(\frac{\Omega_1}{\gamma},\gamma{T})|00\rangle\langle00|\nonumber\\
&&+f(\frac{\Omega_1}{\gamma},\gamma{T})|10\rangle\langle01|+f^{\ast}(\frac{\Omega_1}{\gamma},\gamma{T})|01\rangle\langle10|.
\eeqa In order to quantify the degree of entanglement, we adopt
the concurrence $C$ defined by Wooters \cite{Woo1998}. The
concurrence varies from $C=0$ for an unentangled state to $C=1$
for a maximally entangled state. For two qubits, in the "Standard"
eigenbasis: $|1\rangle\equiv|11\rangle$,
$|2\rangle\equiv|10\rangle$, $|3\rangle\equiv|01\rangle$,
$|4\rangle\equiv|00\rangle$, the concurrence may be calculated
explicitly from the following: \be
C=\max\{\lambda_1-\lambda_2-\lambda_3-\lambda_4,0\}, \ee where the
$\lambda_{i}$($i=1,2,3,4$) are the square roots of the eigenvalues
\textit{in decreasing order of magnitude} of the "spin-flipped"
density matrix operator
$R=\rho_s(\sigma^{y}\otimes\sigma^{y})\rho^{\ast}_s(\sigma^{y}\otimes\sigma^{y})$,
where the asterisk indicates complex conjugation. The concurrence
related to the density matrix $\rho_s$ can be written as \be
C_s=2\max[0,|f(\frac{\Omega_1}{\gamma},\gamma{T})|-\sqrt{a(\frac{\Omega_1}{\gamma},\gamma{T})d(\frac{\Omega_1}{\gamma},\gamma{T})}].
\ee The mutual information of the stationary state of two qubits
is defined by $I(\rho_s)=S(\rho_1)+S(\rho_2)-S(\rho_s)$, where
$S(\rho)=-{\mathrm{Tr}}(\rho\log_2\rho)$ is the Von-Neumann
entropy of $\rho$, and $\rho_1={\mathrm{Tr}}_2\rho_s$ and
$\rho_2={\mathrm{Tr}}_1\rho_s$ are the reduced density operators
of the qubit 1 and the qubit 2 respectively. The mutual
information can used to measure the total correlation of two
qubits. We can easily obtain the analytical expression of the
mutual information $I$ as follows, \beqa
I&=&-(a+b)\log_2(a+b)-(c+d)\log_2(c+d)\nonumber\\
&&-(a+c)\log_2(a+c)-(b+d)\log_2(b+d)\nonumber\\
&&+a\log_2a+d\log_2d\nonumber\\
&&+\beta_{+}\log_2\beta_{+}+\beta_{-}\log_2\beta_{-},\eeqa where
\be \beta_{\pm}=\frac{b+c\pm\sqrt{(b-c)^2+4|f|^2}}{2}. \ee
Firstly, we consider the case in which two qubits are initially in
the Bell state $|\Phi^{-}\rangle$, i.e., the Bell singlet state.
In Fig.1, the stationary state concurrence $C_s$ and the mutual
information $I$ of the stationary state are plotted as the
function of the parameter $\gamma{T}$. It is shown that the values
$\gamma{T}^m_i$ ($i=0,1,2,...,9$ in this case) of the scaled
action time which locally maximize the stationary state
entanglement always locally maximize the mutual information, i.e.
the total correlation of two qubits. Both the mutual information
of the stationary state and the stationary state concurrence
oscillate with $\gamma{T}$, which implies one can control both the
entanglement and mutual information of the stationary state of two
qubits in the collective dephasing environment by adjusting the
scaled action time $\gamma{T}$ of the locally driving field. For
analyzing Fig.1 in details, we orderly label those special points
with $\gamma{T}_i$ ($i=1,2,3,...$) at which the stationary states
transit from the entangled state to the separable state or vice
versa. Namely, $C_s(\gamma{T})>0$ when $0\leq{T}<T_1$ or
$T_{2k}<T<T_{2k+1}$ ($k=1,2,...9$ in Fig.1), and
$C_s(\gamma{T})=0$ when $T_{2k-1}\leq{T}\leq{T}_{2k}$
($k=1,2,...9$ in Fig.1) or $T\geq{T_{19}}$. In those windows of
$T\in[T_{2k-1},T_{2k}]$, the initial bell singlet state eventually
becomes separable in the dynamical evolution governed by the
Eq.(3) though the total correlation of two qubits in the
stationary state is not zero. The length $\gamma(T_{2k}-T_{2k-1})$
($k=1,2,...,9$) of those windows increase with $k$. We can find
that in the case with strong intensity of external field, i.e.
$\Omega_1\gg\gamma$, the stationary state concurrence decreases
from one to zero with the increase of the scaled action time
$\gamma{T}$ from zero to $\gamma{T_1}$. With the further increase
of $\gamma{T}$ from $\gamma{T}_2$ to $\gamma{T^m_1}$, the
stationary state concurrence increases from zero to the local
maximal value $C_s(\gamma{T^m_1})$. The local maximal values
$C_s(\gamma{T^m_i})$ are in decreasing order of magnitude, namely
$C_s(\gamma{T^m_i})>C_s(\gamma{T^m_{i+1}})$. Now we turn to
discuss the mutual information of the stationary state of two
qubits. $I(\gamma{T^m_i})$ are also the local maximal values. When
$T_{2k}\leq{T}\leq{T}_{2k+1}$, the total correlation of two qubits
firstly increases with $\gamma{T}$ and achieves its local maximum
at $\gamma{T^m_k}$, then decreases with $\gamma{T}$. This
dynamical behavior is consistent with the entanglement of two
qubits. When $T_{2k-1}\leq{T}\leq{T}_{2k}$, the stationary state
is separable but the total correlation is not zero and has a peak
value.

\begin{figure}
\centerline{\includegraphics[width=2.5in]{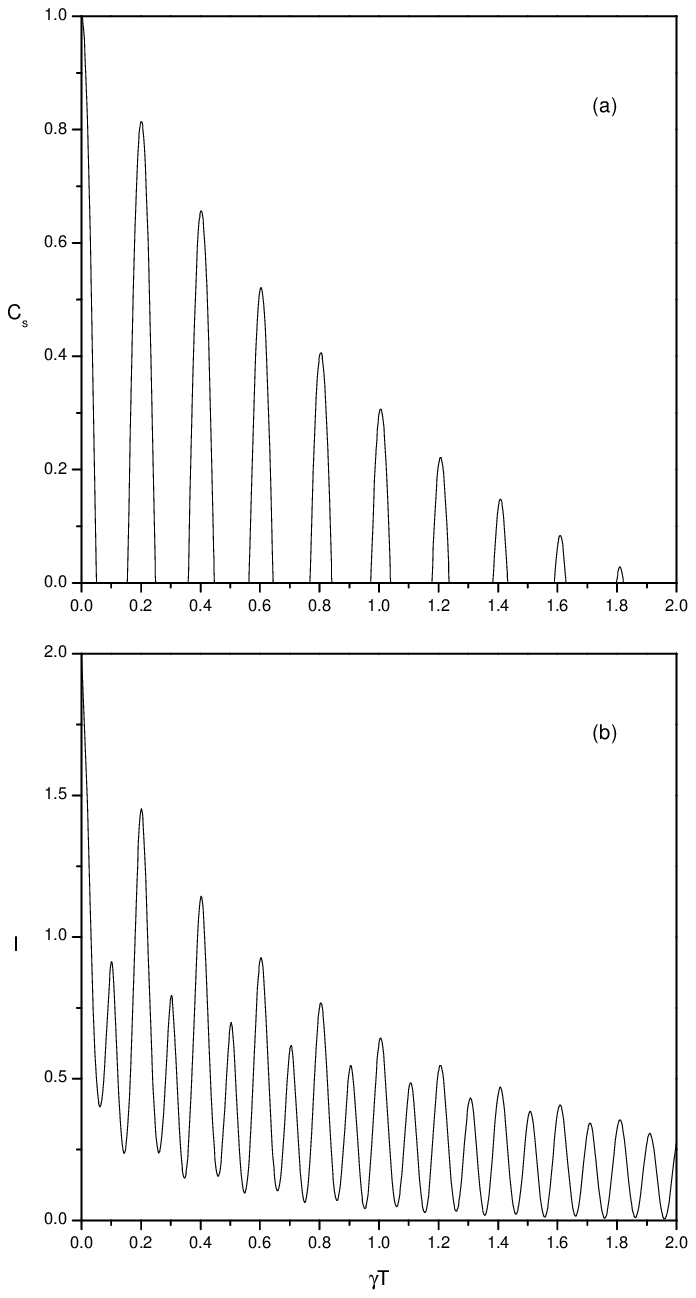}}
\caption{The stationary state concurrence $C_s$ and the mutual
information $I$ of the stationary state are plotted as the
function of the parameter $\gamma{T}$ with
$\Omega_1/\gamma=31.25$. In this case, two qubits are initially in
the Bell state $|\Phi^{-}\rangle$. By comparing the concurrence in
(a) and the mutual information in (b), it is not difficult to
verify that the mutual information is always larger than the
concurrence.}
\end{figure}

In what follows, we consider the case in which two qubits are
initially in the fragile entangled state $|\Psi^{+}\rangle$. If
the external driving field is absence, two qubits lose its
entanglement more rapidly than the disappearance of its coherence.
Certainly, it is a disadvantage for many quantum information
processes such as the solid state quantum computation. Here, we
shall show that a finite-time external driving field can protect
two qubits from completely disentanglement. In Fig.2, we display
the stationary state concurrence and the mutual information of the
stationary state as the function of the parameter $\gamma{T}$.
When $\gamma{T}$ is chosen as a very small value, the stationary
state is still separable and the total correlation is weaken. For
conveniently analyzing the Fig.2, we still orderly label those
special points with $\gamma{T}^{\prime}_i$ ($i=1,2,3,...$) at
which the stationary states transit from the entangled state to
the separable state or vice versa. After carefully comparing the
Fig.2(a) and Fig.1(a), we may observe \be
{T}^{\prime}_i\geq{T}_i,~~~{\mathrm{and}}~~~{T}^{\prime}_i\leq{T}_i.
\ee The above inequalities imply that two qubits initially in Bell
states $|\Phi^{-}\rangle$ or $|\Psi^{+}\rangle$ can not
simultaneously evolve into stationary entangled states in their
evolutions governed by Eq.(3) with fixed parameter $T$. Being
similar with Fig.1, we can also find that in the range of
${T}^{\prime}_{2k-1}\leq{T}\leq{T}^{\prime}_{2k}$, the behaviors
of entanglement and total correlation are consistent. Both the
concurrence and mutual information simultaneously achieve their
local maximal values.

\begin{figure}
\centerline{\includegraphics[width=2.5in]{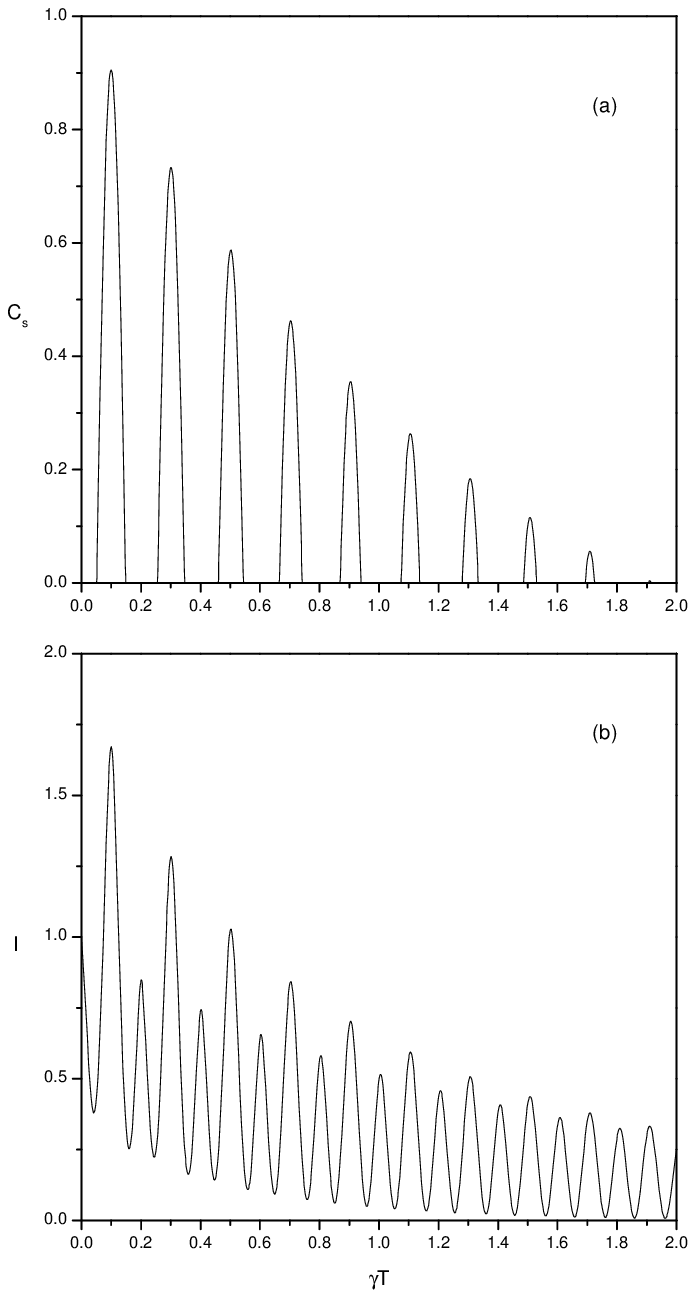}} 
\caption{The stationary state concurrence $C_s$ and the mutual
information $I$ of the stationary state are plotted as the
function of the parameter $\gamma{T}$ with
$\Omega_1/\gamma=31.25$. In this case, two qubits are initially in
the Bell state $|\Psi^{+}\rangle$. }
\end{figure}

\section * {III. TWO QUTRITS IN THE COLLECTIVE DEPHASING CHANNEL}

In this section, we briefly discuss the possible generalization to
two spin 1 $\hat{j}^(1)$ and $\hat{j}^(2)$, i.e. two qutrits.
$\{|-1\rangle,|0\rangle,|1\rangle\}$ are three eigenvectors of the
$z$-component of a spin 1. The collective dephasing of two spin 1
can be described by the master equation \be
\frac{\partial\hat{\rho}}{\partial{t}}=\frac{\gamma}{2}(2\hat{L}_{z}\hat{\rho}\hat{L}_{z}-\hat{L}^2_{z}\hat{\rho}-\hat{\rho}\hat{L}^2_{z}),
\ee where $\gamma$ is the decay rate. $\hat{L}_{z}$ are the
collective spin operator defined by \be
\hat{L}_{z}=\sum^{2}_{i=1}\hat{j}^{(i)}_{z},\ee where $\hat{j}_z$
for each spin 1 is defined by
$\hat{j}_{z}=|1\rangle\langle{1}|-|-1\rangle\langle{-1}|$. By
making use of the well-known Peres-Horodecki criterion
\cite{Peres1996,Horodecki1996}, we can investigate whether the
stationary state is entangled or not. For any initial density
operator $\rho$, the sufficient condition that the stationary
state is entangled can be derived as follows: The equation \be
x^3-\xi{x}^2+\zeta{x}+\eta=0, \ee where \beqa
\xi&=&\rho^2_{1,1;1,1}+\rho^2_{0,0;0,0}+\rho^2_{-1,-1;-1,-1},\nonumber\\
\zeta&=&\rho_{1,1;1,1}\rho_{0,0;0,0}+\rho_{1,1;1,1}\rho_{-1,-1;-1,-1}\nonumber\\
&&+\rho_{0,0;0,0}\rho_{-1,-1;-1,-1}-|\rho_{0,-1;-1,0}|^2\nonumber\\
&&-|\rho_{1,0;0,1}|^2-|\rho_{1,-1;-1,1}|^2,\nonumber\\
\eta&=&-\rho_{1,1;1,1}\rho_{0,0;0,0}\rho_{-1,-1;-1,-1}\nonumber\\
&&+\rho_{1,1;1,1}|\rho_{0,-1;-1,0}|^2+\rho_{-1,-1;-1,-1}|\rho_{1,0;0,1}|^2\nonumber\\
&&+\rho_{0,0;0,0}|\rho_{1,-1;-1,1}|^2\nonumber\\
&&-\rho_{1,-1;-1,1}\rho^{\ast}_{0,-1;-1,0}\rho^{\ast}_{1,0;0,1}\nonumber\\
&&-\rho^{\ast}_{1,-1;-1,1}\rho_{0,-1;-1,0}\rho_{1,0;0,1},\nonumber\\
&&\rho_{i,j;k,l}\equiv\langle{i,j|\rho|k,l\rangle}, \eeqa has a
negative root, or at least one of the following two inequalities
are satisfied \be
|\rho_{1,-1;0,0}|^2>\rho_{1,0;1,0}\rho_{0,-1;0,-1},\ee or \be
|\rho_{0,0;-1,1}|^2>\rho_{0,1;0,1}\rho_{-1,0;-1,0}. \ee

\section * {IV. CONCLUSIONS}

In this paper, we comparatively investigate the mutual information
of the stationary state and the stationary state entanglement
quantified by concurrence of two locally driven qubits in the
presence of collective dephasing. We show that one can transform
the fragile entangled states into the stationary entangled states
under the collective dephasing by making use of a finite-time
external driving field. The local maximal value of the stationary
state concurrence corresponds to the local maximal value of mutual
information, i.e. the total correlation of two qubits. We show how
a finite-time external driving field can control the entanglement
and total correlation of the stationary state of two qubits under
the collective dephasing environment. Moreover, we investigate the
model of two spin 1 in the presence of collective dephasing. Since
the stationary state strongly depends its initial state in this
case, we obtain the general sufficient condition that the
stationary state is entangled. It is very interesting to study how
to protect the fragile entangled state of two qutrit in the
collective dephasing environment. The results will be discussed
elsewhere. In the future work, it may be very interesting to apply
the present results to some realistic quantum information
processes, such as quantum computation based on the quantum dots
or Josephson Junctions.

\section*{ACKNOWLEDGMENT}

This project was supported by the National Natural Science
Foundation of China (Project NO. 10174066).

\bibliographystyle{apsrev}
\bibliography{refmich,refjono}

\end{document}